\documentclass[aps,pre,twocolumn,groupedaddress,showpacs,floatfix]{revtex4}
\usepackage{amsmath}
\usepackage{amssymb}
\usepackage{graphicx}

\begin{document}

\title{What's in a Name?} 

\author{Luciano da Fontoura Costa} 
\affiliation{Institute of Physics at S\~ao Carlos. 
University of S\~ ao Paulo, S\~{a}o Carlos,
SP, PO Box 369, 13560-970, 
phone +55 162 73 9858, FAX +55 162 71
3616, Brazil, luciano@if.sc.usp.br}

\date{20th August 2003}

\begin{abstract}   

Among the several findings deriving from the application of complex
network formalism to the investigation of natural phenomena, the fact
that linguistic constructions follow power laws presents special
interest for its potential implications for psychology and brain
science.  By corresponding to one of the most essentially human
manifestations, such language-related properties suggest that similar
dynamics may also be inherent to the brain areas related to language
and associative memory, and perhaps even consciousness.  The present
work reports a preliminary experimental investigation aimed at
characterizing and modeling the flow of sequentially induced
associations between words from the English language in terms of
complex networks.  The data is produced through a psychophysical
experiment where a word is presented to the subject, who is requested
to associate another word.  Complex network and graph theory formalism
and measurements are applied in order to characterize the experimental
data.  Several interesting results are identified, including the
characterization of attraction basins, association asymmetries,
context biasing, as well as a possible power-law underlying word
associations, which could be explained by the appearance of strange
loops along the hierarchical structure underlying word categories.

\end{abstract}


\maketitle

\begin{verse}
  `$\ldots$ that which we call a rose \\
   By any other name would smell as sweet'  \\
 \emph{(Romeo and Juliet, ACT II)}
\end{verse}

\section{Introduction}

Despite its long tradition in mathematics and computer science, graph
theory \cite{Bollobas:2002} has reached great popularity only recently
through innovative research in the novel area which became known as
complex networks \cite{Albert_Barab:2002}.  By integrating theoretical
principles, especially from statistical mechanics, with experimental
and simulated data, recent investigations have shown that several
important natural phenomena such as infectious diseases, ecological
systems, protein folding, society and the internet, are characterized
by scale-free and/or small-world behavior as far as their connectivity
is concerned \cite{Albert_Barab:2002}.  In particular, studies
modeling linguistics in terms of complex networks have indicated that
several aspects of human language, such as word proximity
\cite{Ferrer:2001} and synonyms \cite{Albert_Barab:2002,Motter:2002},
are at least partially characterized by power law behavior.  As
language corresponds to one of the most essential manifestations of
the human brain, such findings can be taken as an indication that that
complex structure, or at least its portions more closely related to
language and associations, may also be intrinsically organized
according to power laws and scale free behavior \cite{Motter:2002}.
As the conscious and predominantly sequential flow of ideas in humans,
James's \emph{fringe of consciousness} \cite{James:1955}, is closely
related to the externalization of ideas through language, it is also
possible that the scale free properties found in linguistic structures
can also be an intrinsic property of consciousness.

The current work aims at investigating such possibilities through a
psychophysical experiment involving human subjects to associate words
from the English language.  By understanding the presented words and
associations as graph nodes and edges, respectively, it is possible to
perform a quantitative analysis of the digraph connections by
considering statistics (average and standard deviation) of network
measurements such as the node degree, the average length, and the
clustering coefficient.

The current article starts by describing the experimental approach and
proceeds by analysing and discussing the respectively obtained data.

\section{The Psychophysical Experiment}

Along the last decades, psychophysics has establishing itself as an
important area in psychology and neuroscience, providing invaluable
means for quantifying perception.  Provided the experiments are
carefully devised and conducted, objective and relatively precise
information can be obtained about the dynamics of perception.  As in
physics, the experiments have to be planned and performed while most
factors likely to influence the investigated phenomena are kept
constant.  The popularization of personal computers has motivated the
use of such machines for automating of psychophysical experiments,
accounting for enhance repeatability and standardization.

In this work, a program was developed in SCILAB with the specific
finality of investigating associations between words from the English
language.  Starting with the word 'sun', the subject is prompted to
associate a subsequent word.  No specific instructions are given
regarding the type of association, except that special characters,
plurals and verb conjugations are to be completely avoided.  There is
no time limit for providing the new word, and the experiment can be
broken into several sections, while collecting all the obtained data
stored into files.  An illustration of the first steps of the
experiment is provided below, where the words supplied by the subject
are represented in italic:

\begin{trivlist}
\item sun $\mapsto$ \emph{desert}
\item desert $\mapsto$  \emph{pyramid}
\item sun $\mapsto$ \emph{gold}
\item pyramid $\mapsto$ \emph{triangle}
\item pyramid $\mapsto$ \emph{desert}
\item triangle$\mapsto$ \emph{square}
\item $\ldots$
\end{trivlist}

Observe that the only predefined word is that presented first, all the
others being subsequently defined by the subject. The presented and
suggested words are henceforth referred to as \emph{presented word}
and \emph{input word}, respectively. After each new word is input, its
presence in the current list of words is verified, the word being
included otherwise.  Each word is treated as a graph node, and each
pair of words is understood as a graph edge $($presented word, input
word$)$.  The therefore obtained direct graph (i.e. a digraph), with
the frequency of each association treated as the weight of the
respective edge, provides an interesting formal representation of the
word associations. The whole sequence of presented and input words is
recorded for further analysis.  In order to guarantee the words to be
presented in a uniform fashion, in the sense that each word is
presented about as many times as the others, a density probability
function $p(w)$ describing the number of times each word is presented
is kept all times.  The presented words are drawn from the
complemented density function, i.e. $max \{ p(w) \} -p(w)$, so that
the less frequently presented words have higher likelihood to be
chosen, leading to a levelling effect.  The experiment terminates
after a pre-defined number of words are presented, and the more recent
input words, which have consequently been presented only a few times,
are excluded from the data and respective network.

\section{Results}

\begin{figure}
 \begin{center} 
   \includegraphics[scale=.35,angle=0]{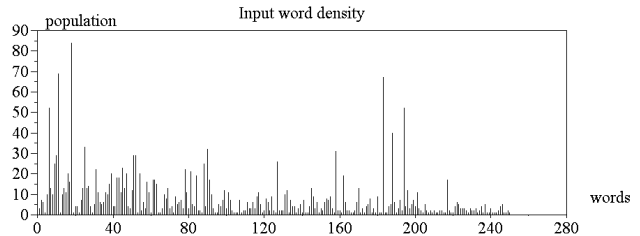}
   \caption{The histogram of input words.~\label{fig:inp_word_dens}}
\end{center}
\end{figure}

\begin{figure}
 \begin{center} 
   \includegraphics[scale=.35,angle=0]{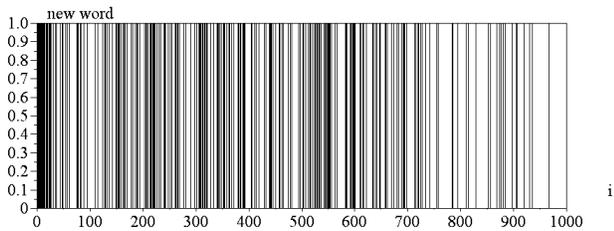}
   \caption{The occurrence of new words along the presentation sequence, indexed by $i$.~\label{fig:new_word}} 
\end{center}
\end{figure}

\begin{figure}
 \begin{center} 
   \includegraphics[scale=.35,angle=0]{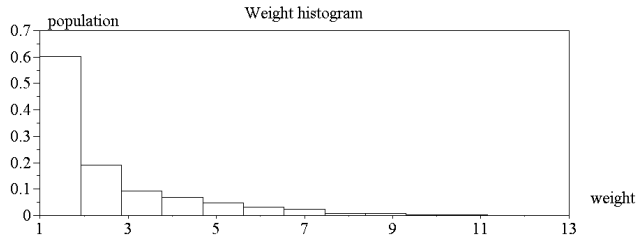}
   \caption{Histogram of weights, i.e. the number of times specific associated pairs of words were produced during the experiment.~\label{fig:weight_hist}} 
\end{center}
\end{figure}

\begin{figure}
 \begin{center} 
   \includegraphics[scale=.35,angle=0]{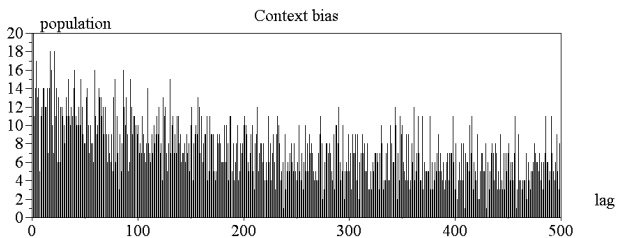}
   \caption{The population of equal input words distant one another by specific lag values.~\label{fig:context}} 
\end{center}
\end{figure}

\begin{figure}
 \begin{center} 
   \includegraphics[scale=.35,angle=0]{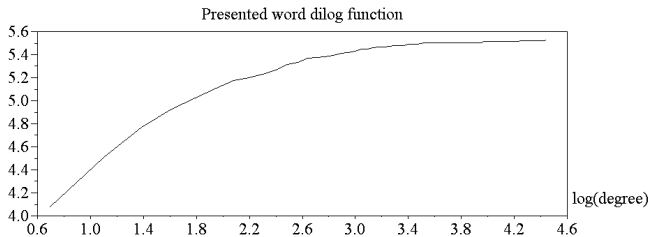}
   \caption{Loglog representation of the cumulative output node 
   degree.~\label{fig:dilog_output}} 
\end{center}
\end{figure}

\begin{figure}
 \begin{center} 
   \includegraphics[scale=.35,angle=0]{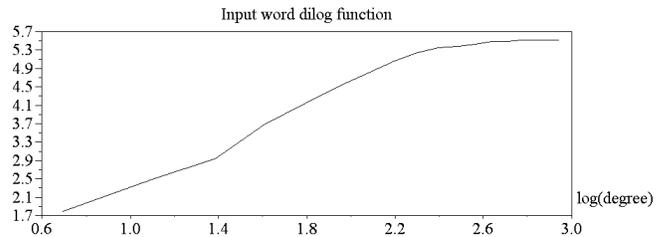}
   \caption{Loglog representation of the cumulative input node
   degree.  An approximatedly straight region is observed along
   the lefthand side of the graph.~\label{fig:dilog_input}} 
\end{center}
\end{figure}

The above experiment was performed with a single subject along a whole
week, totaling 305 different words from which 250 words were chosen
(the remainder, more recently input words, were discarded for the sake
of enhanced uniformity). A total of 1930 associations were recorded.
The types of the input words is given in Table~\ref{tab:finp}, and
Table~\ref{tab:assoc_types} shows the frequency of types of
associations.  Figure~\ref{fig:inp_word_dens} presents the population
of each input word, and Figure~\ref{fig:new_word} depicts the
occurrence of new words along the presentation stages identified by
$i$.  Figure~\ref{fig:weight_hist} gives the histogram of repeated
associations.  The average and standard deviation of the node degree
$k$, clustering coefficient $C$ and average length $\ell$ are
presented in Table~\ref{tab:av_st}.  Figure~\ref{fig:context} shows
the histogram of equal words apart by specific lags along the
presentation sequence.  For instance, the ordinate value at lag 100
indicates the number of equal words distant of lag along the sequence.
For the sake of enhanced uniformity, the sequence is considered up to
its total length minus the maximum lag value.  The loglog curves of the
cumulative output and input node degree (recall that we are dealing
with a digraph) are presented in Figures~\ref{fig:dilog_output}
and~\ref{fig:dilog_input}.

\begin{table}
  \begin{tabular}{||l|r||}  \hline
    noun        &  161 \\   \hline
    adjective   &   62 \\  \hline
    verb        &   15 \\   \hline
    proper noun &    6 \\   \hline
    other       &    6 \\  \hline
    TOTAL       &  250 \\   \hline
  \end{tabular}
\caption{Total of words by category.~\label{tab:finp}}
\end{table}

\begin{table}
  \begin{tabular}{||c|c|c|c|c|c||}  \hline
                & noun  & adjective & verb & proper noun & other \\   \hline
    noun        & 595   &  233      & 62   &     26      &  16   \\   \hline
    adjective   & 612   &  222      & 48   &      9      &   0   \\   \hline
    verb        &  37   &   20      &  7   &      1      &   1   \\   \hline
    proper noun &   9   &    6      &  0   &      7      &   0   \\   \hline
    other       &   7   &    0      &  1   &      0      &  11   \\   \hline
  \end{tabular}
\caption{Number of associations by category.~\label{tab:assoc_types}}
\end{table}

\begin{table}  
  \begin{tabular}{||c|c||}  \hline
    k           & 7.72 $\pm$ 10.93  \\  \hline
    C           &  0.075 $\pm$ 0.17    \\   \hline
    $\ell$      &  3.32 $\pm$ 0.95    \\   \hline
  \end{tabular}
\caption{The node degree $k$, clustering coefficient $C$ and average
length $\ell$ for the network obtained in the psychophysical
experiment.~\label{tab:av_st}}
\end{table}

\section{Discussion}

The several interesting trends and phenomena identified by analysis of
the experimental data are characterized and briefly discussed in the
following:

\emph{A. Attractor formation:} As shown in Figure~\ref{fig:new_word},
the number of new words input by the user tended to diminish, reaching
a near equilibrium state where very few new words are likely to be
added.  This suggests an attraction basin defined by the initial word.

\emph{B. Word density asymetry:} As illustrated in
Figure~\ref{fig:inp_word_dens}, the subject tended to enter some words
more often than others.  Particularly, there was a generalized
preference for adjectives such as \emph{good} and \emph{long}, among
others.  This is hardly surprising, as adjectives are more immediately
applicable to several words.

\emph{C. Edges asymmetry:} As clearly seem from
Table~\ref{tab:assoc_types}, not every association is reciprocal,
i.e. the existence of an edge $(i,j)$ does not necessarily implies the
presence of $(j,i)$.  Examples of such asymmetric cases obtained in the
considered experiment include $(sky,blue)$ and $(blue,sky)$.  To some
extent, such asymmetries are observed in cases involving a more common
word followed by a less common one, such as a general adjective and a
specific noun.  Another characteristic that has been verified from the
experiment is the tendency of the associations to correspond to
synonyms and antonyms, especially regarding pairs of adjectives.

\emph{D. Wide variation of node degree:} The high standard deviation
obtained for the node degree indicates that the number of associations
induced by each presented word varies considerably.  It is possible
that more common words which usually appear connected to several other
words, such as adjectives, tend to favor higher number of
associations. 

\emph{E. Associations asymmetry:} One of the clearest results deriving
from the reported experiment was the fact that some associations
tended to be much more stable than others, in the sense that they were
more systematically repeated and yielded a smaller number of
variations.  Examples of such cases include $(bread,butter)$ and
$(pecker,wood)$.  This property seems to be connected to the node
degree wide dispersion, in the sense that association pairs involving
at least one word characterized by higher node degree tended to favor
a higher number of different associations.

\emph{F. Context biasing:} As is clear from Figure~\ref{fig:context},
the choice of a word by the subject tended to be influenced by those
more recently input.  The memory effect seems to disappear for lags
higher than 250 presentations.

\emph{G. Small-world features} The relatively low average length shown
in Table~\ref{tab:av_st} suggests that the obtained association
follows the small-world paradigm.  This is an immediate consequence of
the fact that the experiment inherently targets word associations.

\emph{H. Power-law features:} As could be expected, the output and
input degree distributions shown in Figures~\ref{fig:dilog_output}
and~\ref{fig:dilog_input} resulted markedly different, with the latter
being more compatible with power-law scaling, especially at the
initial portion of the curve.

\begin{figure}
 \begin{center} 
   \includegraphics[scale=.55,angle=0]{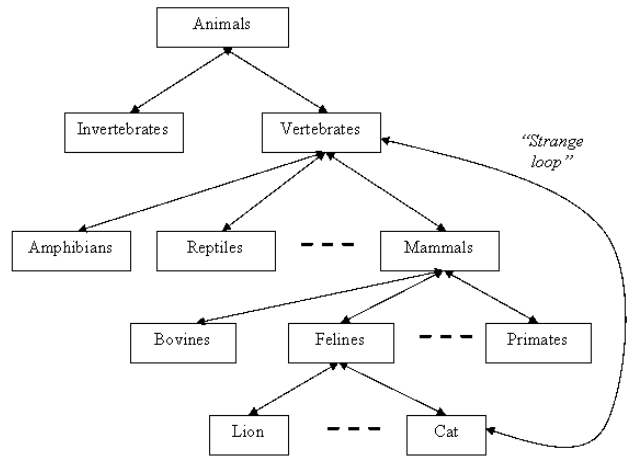}
   \caption{Hierarchical organization of words and formation of
strange loops imply that words in higher hierarchical levels
acquire higher number of associations.~\label{fig:hierarch}} \end{center}
\end{figure}

As the limited power-law trend corresponds to the possibly more
complex and interesting identified features, it is further discussed
in the following, including a possible explanation.  Although
difficult to be defined, the conscious portion of thinking is a
predominantly sequential process.  While solving a problem, or just
relaxing, the flow of ideas and concepts takes place as a sequence of
ideas associated in some way which is highly dependent on the context
defined by the more recent thoughts.  To a large extent, the
successive ideas are characterized by some strong or weak association.
For instance, after thinking about the sky, next possible ideas are
likely to be blue, air, sun or clouds.  Therefore, at least part of
the flow of thoughts can be thought in terms of a Markovian system.
At the same time, memories are often related to associations
\cite{Humphreys:1989}.  From the computational point of view, it is
possible to enhance the storage potential by organizing the stored
concepts in a hierarchical fashion, so that the description of new
concepts at lower hierarchical levels can include only the features
not covered by the upper levels.  For instance, the description of a
cat can be derived from that of mammals, including only those
characteristics that are intrinsic to cats (see
Figure~\ref{fig:hierarch}).  This concept of inheritance leads
naturally to associations between concepts and ideas, even between two
non-adjacent hierarchical levels, a phenomenon that can be related to
Hofstadter 'strange loops' \cite{Hofstadter:1999}.  Though additional
features are certainly incorporated into the brain dynamics, such
hierarchical and associative schemes lead to the interesting situation
where several concepts end up associated, even if indirectly, to those
in the upper hierarchies.  Consequently, the concepts tend to become
more and more associated as one moves from the lower to the upper
hierarchical levels, possibly leading to a rich gets richer scheme,
and hence to scale free organization.

Given that the adjacency matrix of the obtained graph can be
immediately understood as the transition matrix of a Markovian
systems, it is possible to use Monte Carlo simulation in order to
produce sequences of associated words, such as that illustrated below.
As the context is limited to one association level, such sequences are
characterized by subsequent repetitions of words.

\begin{verse}
horse, brown, bear, brown, sugar, sweet, \\
good, earth, land, good, well, good, time, \\
out, sun, hot, water, cold, water, cold, \\
wool, sheep, four, clock, six, tea, leaf,\\
thin,  sheet, wide, field
\end{verse}

\section{Concluding Remarks}

The present work has illustrated how complex network and graph theory
concepts and formalisms can be applied to characterize human cognitive
activities, namely the association of words.  While previous related
works such as Motter et al.\cite{Motter:2002} investigated word
associations through the use of static databases, the current approach
considered psychophysical experiments.  The main differences implied
by such an approach are the fact that the importance of associations
can be inferred from the respective frequencies.  In addition, a
random element is implied by the fact that the user is likely to vary
the chosen associations while affected by the context established by
the presentation sequence.

Although limited to a single subject, the obtained experimental
results led to a series of interesting findings, including the
identification of attraction basins, context biasing, association
asymmetries, small-world features, and near power-law scaling of the
node degree.  A putative model possibly underlying the latter
phenomenon, involving the appearance of strange loops in the
hierarchical categorization of words, has also been proposed.  While
extensive additional investigations are required in order to confirm
such preliminary results, it is felt that the identified phenomena are
likely to provide a reasonably formal scaffolding for further
investigating and understanding word associations by humans and even
more sophisticated brain dynamics \cite{Motter:2002}.

Several are the possibilities implied by the reported developments.
First, it is important to note that the specific measurements
extracted from digraphs obtained from different subjects can be
possibly correlated to individual features or even for diagnosis.  At
the same time, it is likely that the obtained graphs will present a
core shared by several subjects, corresponding to those more
established and invariant collective concepts, while the graph
difference residuals could provide interesting information about
intrinsic individual features and preferences.  Another interesting
task would be to extend the reported approach in order to investigate
associations in visual language, for instance by using eye-tracking
systems.  Several possibilities for further investigation can be
defined by considering modified versions of the adopted psychophysical
experiment. For instance, it would be interesting to study situations
where the subject is allowed to enter a continuous flow of associated
words, without any interference from the computer, except the
presentation of the first word.  Although more complex, given the
additional degrees of freedom, such investigations could provide
additional insights about long time memory effects, which are poised
to reduce the number of word repetitions in respective Monte Carlo
simulations.  It would be interesting to compare how such extended
context modifies the properties of the respectively obtained networks.

\begin{acknowledgments}
The author is grateful to FAPESP (processes 99/12765-2 and 96/05497-3)
and CNPq for financial support.
\end{acknowledgments}

 
\bibliography{brainassoc}

\end{document}